\documentclass[conference]{IEEEtran}
\IEEEoverridecommandlockouts
\usepackage{cite}
\usepackage{amsmath,amssymb,amsfonts}
\usepackage{graphicx}
\usepackage{textcomp}
\usepackage{xcolor}
\usepackage{eso-pic}
\def\BibTeX{{\rm B\kern-.05em{\sc i\kern-.025em b}\kern-.08em
    T\kern-.1667em\lower.7ex\hbox{E}\kern-.125emX}}

\begin{document}
\title{Self-Learning Surprise Detection in Spatial Audio for Viewport-Aware 360-Video Streaming}
\author{
\IEEEauthorblockN{Arman Nik Khah}
\IEEEauthorblockA{\textit{Department of Computer Science} \\
\textit{The University of Texas at Dallas}\\
Richardson, TX, USA \\
arman.nikkhah@utdallas.edu}
\and
\IEEEauthorblockN{Ravi Prakash}
\IEEEauthorblockA{\textit{Department of Computer Science} \\
\textit{The University of Texas at Dallas}\\
Richardson, TX, USA \\
ravip@utdallas.edu}
}
\AddToShipoutPictureFG*{\AtPageLowerLeft{\put(\LenToUnit{0.08\paperwidth},\LenToUnit{0.25in}){\parbox{0.84\paperwidth}{\centering\scriptsize Accepted at the 2026 IEEE International Conference on Multimedia and Expo (ICME).\\
\copyright\ 2026 IEEE. Personal use of this material is permitted. Permission from IEEE must be obtained for all other uses, in any current or future media, including reprinting/republishing this material for advertising or promotional purposes, creating new collective works, for resale or redistribution to servers or lists, or reuse of any copyrighted component of this work in other works.}}}}
\maketitle
\begin{abstract}
Adaptive streaming of 360-degree video relies on viewport prediction to allocate bandwidth efficiently.
Current approaches predominantly use visual saliency or historical gaze patterns, neglecting the role of spatial audio in guiding user attention.
This paper presents a self-learning framework for detecting ``surprising'' auditory events---moments that deviate from learned temporal expectations---and demonstrates their utility for viewport prediction.
The proposed architecture combines SE(3)-equivariant graph neural networks with recurrent temporal modeling, trained via a dual self-supervised objective.
A key feature is the natural modeling of \emph{temporal attention decay}: surprise is high at event onset but diminishes as the listener adapts.
Experiments on the AVTrack360 dataset show that integrating audio surprise with visual cues reduces bitrate waste by up to 18\% compared to visual-only methods.
\end{abstract}
\section{Introduction}
\label{sec:introduction}
Unlike traditional video, where the entire frame is always visible, 360-video covers the full sphere around the viewer, yet only a small portion---the \emph{viewport}---is seen at any moment.
Transmitting the entire sphere at high quality is prohibitively expensive.
Modern streaming systems employ \emph{viewport prediction} algorithms that anticipate where the user will look in the near future \cite{yaqoob2020survey360}.
By pre-fetching high-quality tiles for predicted regions, bandwidth can be allocated more efficiently, reducing both network load and perceived latency. 
The accuracy of viewport prediction thus directly impacts both quality of experience (QoE) and system efficiency.

\subsection{The Overlooked Role of Audio}

The vast majority of viewport prediction methods rely exclusively on \emph{visual} cues: image saliency maps, object detection, motion vectors, or historical gaze trajectories \cite{wang2021revisiting}.
In reality, sudden sounds frequently \emph{precede} and \emph{trigger} visual attention shifts.
A honking car, a slamming door, or an unexpected voice from behind the viewer are all examples of auditory events that can cause rapid head rotation toward the sound source.
When such events occur outside the current field of view, visual-only models have no signal to predict the impending gaze shift.
The user turns toward the sound, the streaming system is caught off-guard, and quality degrades precisely when attention is highest.
This represents a significant gap in current viewport prediction research.

\subsection{Challenge: Defining ``Surprise'' in Audio}

Human perception of audio salience is \emph{context-dependent} and \emph{temporal}:
\begin{itemize}
    \item A sustained siren, though loud, is less attention-capturing than its abrupt onset.
    \item A whisper in a silent library is more surprising than a shout at a concert.
    \item An unexpected sound from a new direction draws more attention than a repeated sound from a known source.
\end{itemize}
These observations suggest that audio salience is related to a \emph{deviation from expectation} that decays over time as the listener adapts.
Modeling this \emph{temporality in attention} is essential for accurate viewport prediction.
\subsection{Proposed Solution}
This paper introduces a self-learning framework for detecting surprising auditory events in spatial (First-Order Ambisonics) audio.
The system learns to predict future audio states from past context using a self-supervised objective.
At inference time, \emph{surprise} is defined as the prediction error: events that deviate from learned expectations produce high error, indicating salience.
The architecture is designed with three key principles. (1) \textbf{Geometric Fidelity:} Spatial audio encodes direction via phase relationships between channels. The proposed method projects audio onto a spherical graph and processes it with an SE(3)-equivariant neural network, preserving rotational symmetry. (2) \textbf{Temporal Context:} A recurrent network (GRU) accumulates past information, enabling the model to distinguish a new event from a continuation of an existing one. (3) \textbf{Self-Supervision:} The model requires no labeled ``surprising'' events. It learns the statistics of normal audio and flags deviations, enabling training on unlabeled in-the-wild data.

\subsection{Contributions}
The main contributions of this work are as follows:
\begin{itemize}
    \item \textbf{Novel Spatial Audio Surprise Detection Architecture:} A pipeline combining spherical projection, SE(3)-equivariant encoding, and recurrent temporal modeling for self-supervised surprise detection in First-Order Ambisonics audio.
    \item \textbf{Temporal Attention Decay:} 
    Surprise is naturally high at event onset and decays as the recurrent model adapts, mirroring human attention dynamics.
    \item \textbf{Demonstrated Bitrate Reduction:} Experimental results showing that incorporating audio surprise into viewport prediction reduces bitrate waste in 360-video streaming compared to visual-only baselines.
\end{itemize}
The remainder of this paper is organized as follows.
Section~\ref{sec:related_work} reviews related work in saliency detection, spatial audio analysis, and viewport prediction.
Section~\ref{sec:methodology} describes the proposed methodology in detail.
Section~\ref{sec:experiments} presents experimental setup and results.
Section~\ref{sec:conclusion} concludes with a discussion of limitations and future directions.
\section{Related Work}
\label{sec:related_work}
This section reviews prior work in three areas relevant to the proposed system: visual saliency detection, spatial audio analysis, and viewport prediction for immersive media.

\textbf{Visual Saliency Detection:} Saliency detection aims to identify regions of an image or video that attract human attention.
Traditional methods compute low-level features such as color contrast, edge density, and motion magnitude \cite{itti1998model}.
With the advent of deep learning, convolutional neural networks have been trained on eye-tracking datasets to predict fixation density maps directly from pixel inputs \cite{pan2017salgan}.
Equirectangular projections, commonly used to store 360-video, introduce severe distortions near the poles, biasing saliency predictions toward the equator \cite{coors2018spherenet}.
Recent work addresses this by projecting frames onto polyhedral representations (e.g., cube maps) or by designing spherical convolution operators that respect the underlying geometry \cite{cohen2018spherical}.
Despite these advances, visual saliency methods can only detect salient \emph{visible} content.
An attention-capturing audio event occurring behind the viewer (outside the current viewport) produces no visual signal until after the user has turned.
This work addresses this limitation by incorporating spatial audio, which provides a full-sphere signal regardless of viewing direction.

\textbf{Spatial Audio Analysis:} Spatial audio formats such as First-Order Ambisonics (FOA) encode the full 3D sound field using four channels
\cite{gerzon1973periphony}.
This representation enables sound source localization and immersive playback in VR applications.
Traditional approaches to spatial audio analysis rely on signal processing techniques such as beamforming, time-difference-of-arrival (TDOA) estimation, and spectral decomposition \cite{vanveen1988beamforming}.
Recently, deep learning methods have been applied to tasks such as sound event detection and localization (SELD), where networks are trained to predict the class and direction of active sound sources \cite{politis2020overview}.
The STARSS23 dataset \cite{shimada2023starss23} provides FOA audio with spatial annotations for benchmarking SELD systems.
A key distinction of the proposed method is its focus on \emph{surprise} rather than classification.
Existing SELD methods require labeled training data specifying which sounds are ``interesting.''
In contrast, the proposed self-supervised approach learns the statistics of normal audio and flags deviations, avoiding the need for subjective labeling of salience.
For evaluation of the complete viewport prediction pipeline, we utilize the AVTrack360 dataset \cite{romero2022avtrack360}, which provides synchronized head-tracking logs for 360-degree videos (FOA audio from the original uploads).

\textbf{Viewport Prediction:} Viewport prediction algorithms estimate the user's future gaze direction to enable proactive tile prefetching \cite{rossi2020spherical}.
Methods range from Kalman filter extrapolation \cite{qian2016optimizing} to deep sequence models (LSTMs, Transformers) incorporating visual saliency \cite{xu2020optimizing, chao2021vpt360, nguyen2018content}.
Visual saliency is effective for content-aware prediction, but cannot anticipate off-screen events.
The few works exploring audio cues for viewport prediction \cite{jeong2018viewport} rely on heuristics or supervised labels; the proposed method instead provides a learned, temporally-aware surprise signal that integrates naturally with existing visual pipelines.
\section{Methodology}
\label{sec:methodology}
The proposed solution is designed to learn the statistics of ``normal'' audio sequences without labeled examples,
and subsequently identify ``surprising'' moments---defined as deviations from learned expectations---that
may correlate with attention-capturing events in 360-video streaming.
\subsection{System Overview}
\label{sec:overview}
\begin{figure*}[t]
\centering
\includegraphics[width=0.6\linewidth]{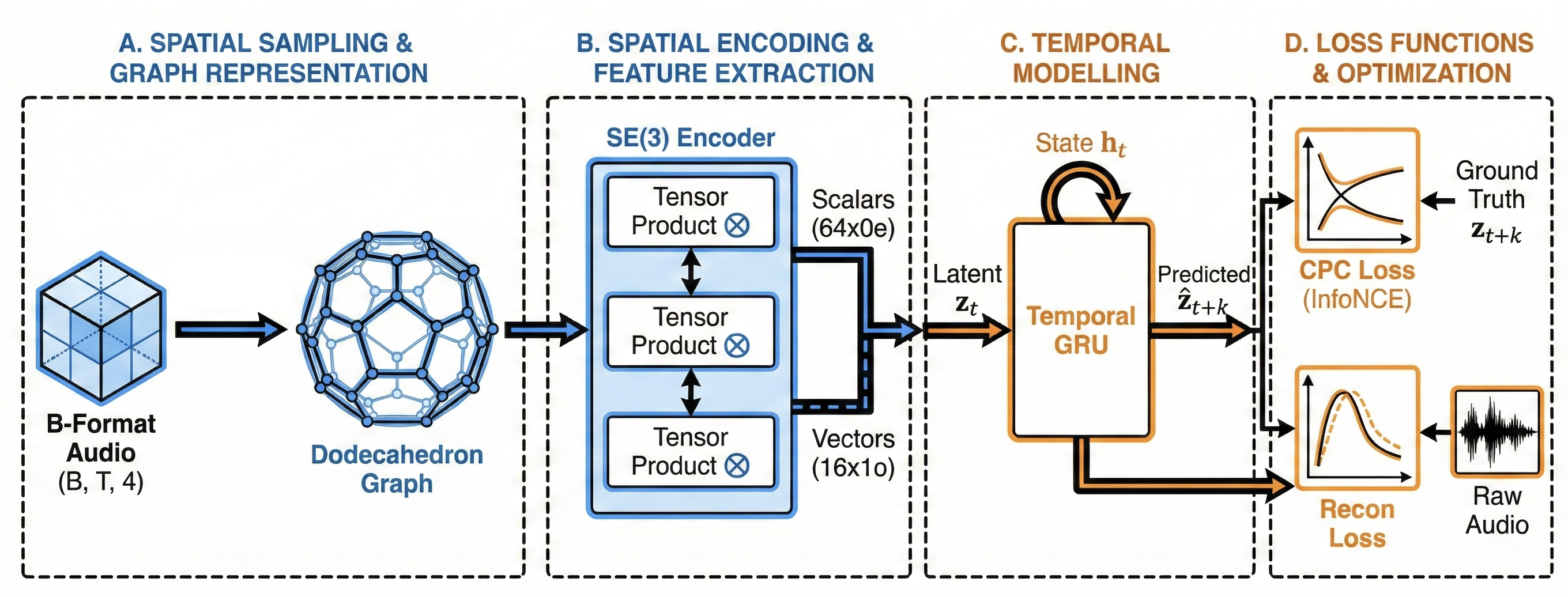}
\caption{\textbf{The proposed Self-Learning Surprise Detection architecture.} (A) \textbf{Spherical Projection:} Ambisonics audio is projected onto a dodecahedral graph. (B) \textbf{SE(3)-Equivariant Encoder:} A graph neural network extracts rotation-invariant scalars ($0e$) and equivariant vectors ($1o$). (C) \textbf{Temporal Modeling:} A GRU predicts future latent states $\hat{\mathbf{z}}_{t+k}$ from history $\mathbf{h}_t$. (D) \textbf{Dual Loss:} The model minimizes Contrastive (CPC) and Reconstruction losses simultaneously.}
\label{fig:arch}
\end{figure*}
The proposed pipeline consists of four stages: (1) \textbf{Spherical Projection:} First-Order Ambisonics (FOA) audio is projected onto a discrete spherical graph. (2) \textbf{Equivariant Encoding:} A rotation-equivariant graph neural network extracts spatial audio features. (3) \textbf{Temporal Modeling:} A Gated Recurrent Unit (GRU) learns temporal dynamics from the encoded features. (4) \textbf{Self-Supervised Training:} A dual-objective loss function trains the model to predict future states. (Figure \ref{fig:arch})

At inference time, the model computes \emph{surprise} scores by measuring the prediction error at each timestep.
A high error indicates that the audio deviated from the model's learned expectations, signaling a potentially salient event.
\subsection{Spherical Projection of Ambisonics Audio}
\label{sec:projection}
Spatial audio in the First-Order Ambisonics (FOA) format is represented by four channels: $W$, $X$, $Y$, and $Z$ \cite{gerzon1973periphony}.
The $W$ channel captures omnidirectional pressure, while $X$, $Y$, and $Z$ encode directional information as
figure-eight patterns aligned with the Cartesian axes.
To avoid geometric artifacts of equirectangular projection, the proposed method samples the audio signal at discrete points on a three-dimensional sphere.
Specifically, a \emph{Dodecahedron} is inscribed in a unit sphere, and its 20 vertices serve as virtual microphone positions.
The Dodecahedron provides near-uniform coverage of the sphere without the singularities inherent in latitude-longitude grids.
Its vertices are defined using the golden ratio $\phi = (1 + \sqrt{5})/2$, yielding 20 points with regular spacing: $(\pm 1, \pm 1, \pm 1)$ and $(0, \pm \phi^{-1}, \pm \phi)$ (cyclic), normalized to the unit sphere.

Each graph node $i$, located at position $\mathbf{p}_i = [p_x^{(i)}, p_y^{(i)}, p_z^{(i)}]^\top$, is assigned a four-dimensional feature vector $\mathbf{f}_i(t)$ derived from the global Ambisonics frame $\mathbf{a}(t) = [W(t), X(t), Y(t), Z(t)]^\top$.
The assignment weights each directional channel by the node's position, yielding a direction-sensitive local representation:
\begin{equation}
    \mathbf{f}_i(t) = \left[ W(t),\; p_x^{(i)} \cdot X(t),\; p_y^{(i)} \cdot Y(t),\; p_z^{(i)} \cdot Z(t) \right]^\top
    \label{eq:node_features}
\end{equation}
This formulation preserves the full 4-channel richness of B-Format audio at each node while introducing spatial locality: nodes facing toward a sound source receive stronger directional components than those facing away.
The result is a set of 20 four-dimensional feature vectors per time step, representing the spatially modulated audio field sampled uniformly on a sphere.

To enable message passing between nodes, a graph structure is constructed by connecting neighboring vertices.
Two vertices are considered neighbors if their Euclidean distance is below a threshold (empirically set to correspond to the dodecahedral edge length),
resulting in a sparse graph with exactly 30 edges for 20 nodes.
For each edge $(i, j)$, a three-dimensional displacement vector $\mathbf{e}_{ij} = \mathbf{p}_i - \mathbf{p}_j$
is computed and stored for use in subsequent processing layers.
\begin{figure}[t]
\centering
\includegraphics[width=0.6\linewidth]{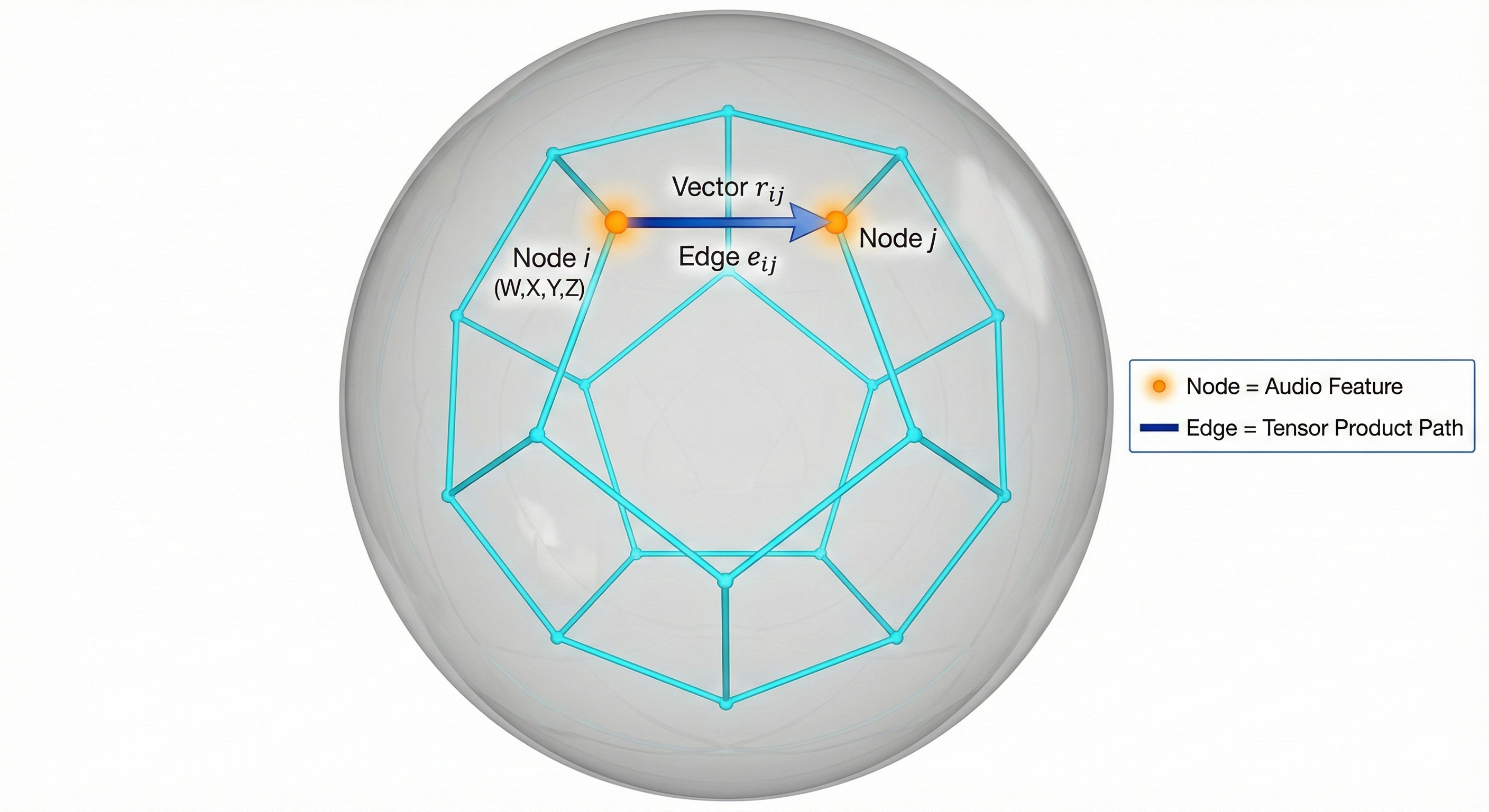}
\caption{\textbf{Geometric Graph Construction.} The Ambisonics audio field is sampled at the 20 vertices of an inscribed dodecahedron.}
\label{fig:dodecahedron}
\end{figure}
\subsection{SE(3)-Equivariant Spatial Encoding}
\label{sec:encoder}
The graph shown in Fig.~\ref{fig:dodecahedron} is processed by a neural network that respects the rotational symmetry of the underlying 3D space.
\subsubsection{The Need for Equivariance}
A standard neural network, when presented with a rotated version of the same audio scene, may produce entirely different outputs.
For spatial audio analysis, this is undesirable: a sound source moving from left to right should be recognized as such,
regardless of the listener's initial orientation.
To address this, the proposed encoder is designed to be \emph{SE(3)-equivariant}:
if the input is rotated by a rotation matrix $R \in SO(3)$, the output transforms in a predictable, mathematically consistent manner.
\subsubsection{Irreducible Representations}
The principle of equivariance is implemented using the theory of \emph{irreducible representations} (irreps) from group theory \cite{thomas2018tensor}.
Features are classified into types based on their transformation properties:
\begin{itemize}
    \item \textbf{Scalars ($\ell=0$):} Values that remain unchanged under rotation (e.g., loudness, pitch).
    \item \textbf{Vectors ($\ell=1$):} Three-dimensional quantities that rotate with the coordinate frame (e.g., sound source direction).
\end{itemize}
By explicitly encoding features as a combination of scalars and vectors, the network can separately reason about
``what'' a sound is (invariant) and ``where'' it is coming from (equivariant).
\subsubsection{Tensor Product Layers}
The core operation in the encoder is the \emph{tensor product}, which combines node features with edge displacement vectors
to propagate information across the graph while preserving equivariance.
Mathematically, if node $i$ has features $\mathbf{f}_i$ and the edge to node $j$ is described by $\mathbf{e}_{ij}$,
the message passed along this edge is:
\begin{equation}
    \mathbf{m}_{ij} = \mathbf{W} \otimes (\mathbf{f}_j \otimes Y(\mathbf{e}_{ij}))
    \label{eq:tensor_product}
\end{equation}
where $Y(\cdot)$ denotes the spherical harmonic decomposition of the edge direction and $\mathbf{W}$ is a learnable weight matrix.
Two tensor product layers are applied in sequence, with gated nonlinearities between them to introduce nonlinearity while preserving equivariance.
The final output of the encoder consists of 64 scalar features and 16 three-dimensional vectors, yielding a 112-dimensional representation per node.
These node-level features are then pooled across the graph via mean aggregation, producing a single 112-dimensional embedding per time step.

\subsection{Temporal Dynamics Modeling}
\label{sec:temporal}
Spatial audio is inherently temporal: a door slamming is surprising not because of its loudness,
but because it deviates from a preceding context of silence.
To capture such dependencies, the sequence of encoded frames is passed through a Gated Recurrent Unit (GRU).
The GRU maintains a hidden state that summarizes past observations: $\mathbf{h}_t = \text{GRU}(\mathbf{z}_t, \mathbf{h}_{t-1})$, where $\mathbf{z}_t$ is the encoded spatial feature at time $t$ and $\mathbf{h}_t$ is the updated context.
This context serves as the basis for predicting future audio states.
\subsubsection{Temporal Compression}
Raw audio at 24~kHz is too dense for efficient recurrent processing.
A \emph{fixed orthogonal projection} (random orthonormal weights, frozen after initialization) downsamples by a factor of 100 to 240 frames per second.
The orthogonal matrix approximately preserves Euclidean distances (Johnson-Lindenstrauss), providing stable, information-preserving compression without the fragility of a learned bottleneck.
\subsection{Self-Supervised Training Objectives}
\label{sec:training}
The model is trained without labeled examples using a combination of two self-supervised objectives:
Contrastive Predictive Coding (CPC) and Reconstruction.
Both objectives are designed to force the model to learn meaningful temporal structure.
\subsubsection{Contrastive Predictive Coding (CPC)}
CPC \cite{oord2018representation} trains the model to predict future representations from past context.
Given context $\mathbf{h}_t$ at time $t$, a projection head predicts the latent $\hat{\mathbf{z}}_{t+k}$ for a future time $t+k$.
The InfoNCE loss encourages similarity between $\hat{\mathbf{z}}_{t+k}$ and the actual future $\mathbf{z}_{t+k}$,
while pushing apart predictions and randomly sampled negatives from the same batch:
\begin{equation}
    \mathcal{L}_{\text{CPC}} = - \sum_{k=1}^{K} \log \frac{\exp(\hat{\mathbf{z}}_{t+k}^\top \mathbf{z}_{t+k} / \tau)}
    {\sum_{j} \exp(\hat{\mathbf{z}}_{t+k}^\top \mathbf{z}_j / \tau)}
    \label{eq:cpc}
\end{equation}
where $\tau$ is a temperature hyperparameter (set to 0.5) and the sum in the denominator runs over all samples in the batch.
This loss forces the model to extract features that are predictive of future dynamics.
\subsubsection{Reconstruction Loss}
To ensure that the latent representation retains low-level signal information, a reconstruction head predicts the original (compressed) audio signal from the GRU context:
\begin{equation}
    \mathcal{L}_{\text{Recon}} = \| \text{ReconHead}(\mathbf{h}_t) - \tilde{\mathbf{a}}_{t+k} \|_2^2
    \label{eq:recon}
\end{equation}
where $\tilde{\mathbf{a}}_{t+k}$ is the target audio after fixed orthogonal projection.
This objective prevents the encoder from discarding perceptually relevant details.
\subsubsection{Loss Balancing and Training Curriculum}
The two losses operate at different scales: CPC loss magnitude is approximately 6.5 (due to its logarithmic formulation),
while reconstruction MSE is approximately 0.3.
To balance gradient contributions, the training follows a two-phase curriculum:
\begin{itemize}
    \item \textbf{Phase 1 (Epochs 1--6):} Pure reconstruction ($w_{\text{Recon}}=1.0$, $w_{\text{CPC}}=0.0$). This ``signal grounding'' phase ensures the encoder learns to preserve audio structure before attempting prediction.
    \item \textbf{Phase 2 (Epochs 7+):} Balanced training ($w_{\text{Recon}}=5.0$, $w_{\text{CPC}}=0.2$). The weights are chosen to equalize gradient magnitudes from both objectives.
\end{itemize}
\subsection{Surprise Computation}
\label{sec:surprise}
At inference time, surprise is computed as the \emph{prediction error} between the model's expected future and the observed future.
For each time step $t$, the GRU context $\mathbf{h}_t$ is used to predict $\hat{\mathbf{z}}_{t+1}$,
and surprise is quantified as the mean squared error against the actual encoding $\mathbf{z}_{t+1}$:
\begin{equation}
    S(t) = \frac{1}{D} \| \hat{\mathbf{z}}_{t+1} - \mathbf{z}_{t+1} \|_2^2
    \label{eq:surprise}
\end{equation}
where $D$ is the latent dimensionality.
A high $S(t)$ indicates that what occurred at time $t+1$ was unexpected given the preceding context---precisely the definition of a ``salient'' or ``surprising'' auditory event.
\subsubsection{Temporality in Attention}
\begin{figure}[t]
\centering
\includegraphics[width=0.7\linewidth]{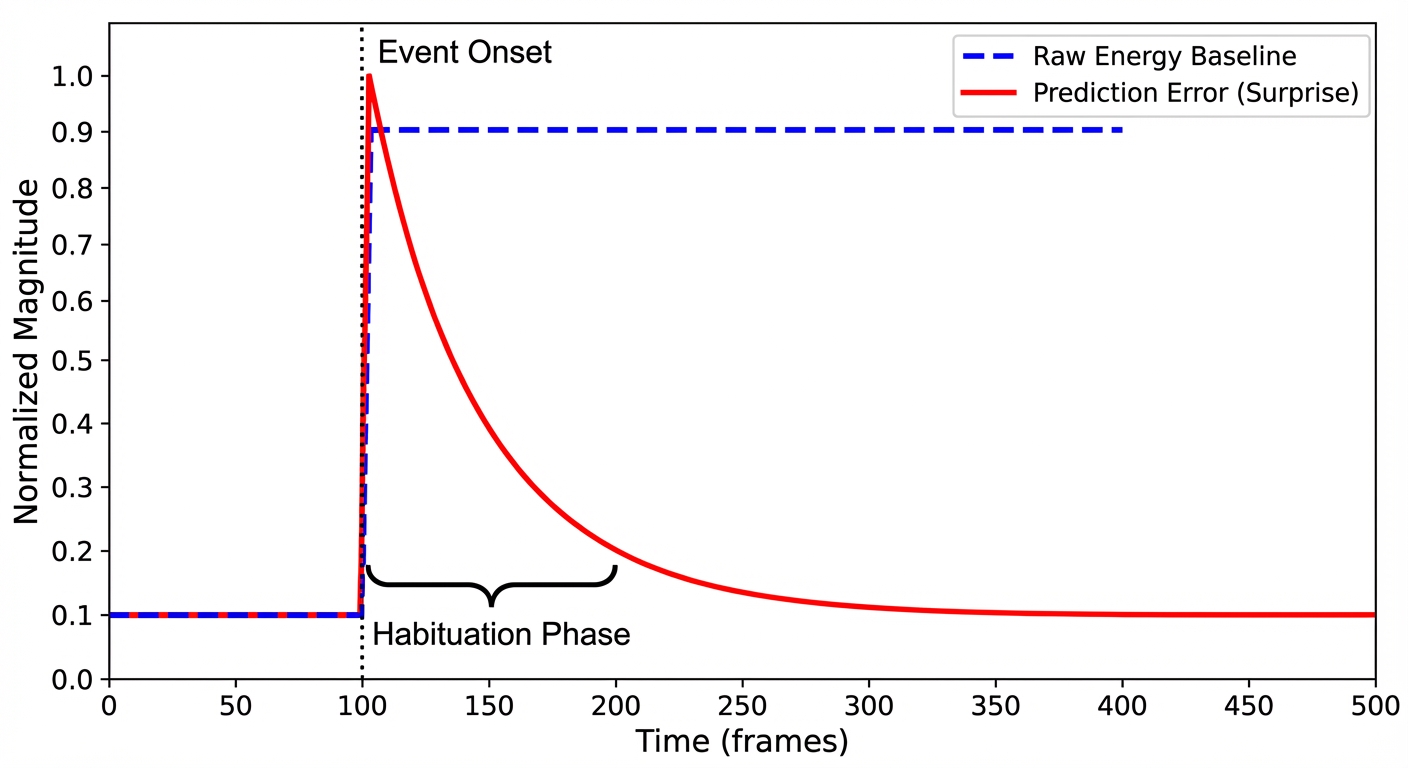}
\caption{\textbf{Illustration of temporal dynamics of Surprise vs. Loudness.} A sustained loud event (e.g., a siren) begins at $t=100$. (Blue) Raw loudness remains high for the duration. (Red) The proposed Surprise score spikes at onset ($t=100$) but decays exponentially as the GRU context $\mathbf{h}_t$ adapts to the new signal, reflecting human habituation.}
\label{fig:habituation}
\end{figure}

The proposed formulation naturally captures \emph{temporal decay} of surprise.
A sudden event (e.g., a door slam) produces a large prediction error at the moment it occurs.
However, once the event enters the GRU's context, it becomes part of the model's expectation.
Subsequent frames, even if loud, are no longer surprising because the model has adapted.
This decay mirrors human attention: a sustained siren is less attention-capturing than its abrupt onset.
As illustrated in Fig.~\ref{fig:habituation}, a raw energy (loudness) baseline remains elevated throughout a sustained event, whereas the proposed surprise score peaks at onset and decays---illustrating that energy alone is insufficient to model attention habituation.
\subsection{Application to Viewport Prediction}
\label{sec:viewport}
The computed surprise scores $S(t)$ can be mapped to spherical coordinates using the vector features from the SE(3)-equivariant encoder.
For viewport prediction in 360-video streaming, the hypothesis is:
\emph{surprising audio occurring outside the current field of view will attract user attention}.
By identifying these moments in advance, the streaming system can prefetch high-quality tiles
for regions likely to be viewed, reducing bandwidth waste and improving perceived quality of experience.
\section{Experimental Results}
\label{sec:experiments}
We aim to demonstrate that while traditional methods perform adequately during normal viewing, they fail catastrophically during ``surprising'' auditory events---precisely the moments that degrade Quality of Experience (QoE) the most.
\subsection{Experimental Setup}
\label{subsec:setup}
We utilized the \textbf{AVTrack360 Dataset} \cite{romero2022avtrack360}, which contains synchronized head-tracking logs for 360-degree videos (FOA audio from the original uploads).
To benchmark the proposed method, we compare three viewport prediction strategies:
\begin{enumerate}
    \item \textbf{LSTM Trajectory Predictor (Baseline):} \cite{fan2017fixation}. A Recurrent Neural Network (RNN) using Long Short-Term Memory units to model non-linear head movement dynamics. This model captures temporal dependencies in the user's velocity and acceleration history to forecast future viewing angles.
    
    \item \textbf{Visual Saliency (Content-Aware):} A spherical Vision Transformer that generates saliency maps based on pixel-level features.\cite{b9}
    
    \item \textbf{Proposed (Audio-Visual Hybrid):} The Visual Saliency model augmented with our \emph{Audio Surprise Signal}.
    The final prediction is a weighted fusion: $\vec{P}_{\text{final}}(t) = (1 - \alpha_t) \vec{P}_{\text{visual}}(t) + \alpha_t \vec{P}_{\text{audio}}(t)$.
    The mixing weight $\alpha_t \in [0, 1]$ is derived from the surprise score $S(t)$ via a learnable sigmoid gating function: $\alpha_t = \sigma(\lambda S(t) + \beta)$, where parameters $\lambda$ and $\beta$ are tuned on a held-out validation set.
    When $S(t)$ is low, $\alpha_t \rightarrow 0$ and the system relies on vision; when surprise spikes, $\alpha_t \rightarrow 1$, shifting attention toward the audio source direction.
    Ablations over alternative fusion strategies (e.g., multiplicative gating, attention-based fusion) showed comparable trends; the sigmoid gate was retained for its low overhead and stable convergence.
\end{enumerate}
\begin{figure}[t]
\centering
\includegraphics[width=0.7\linewidth]{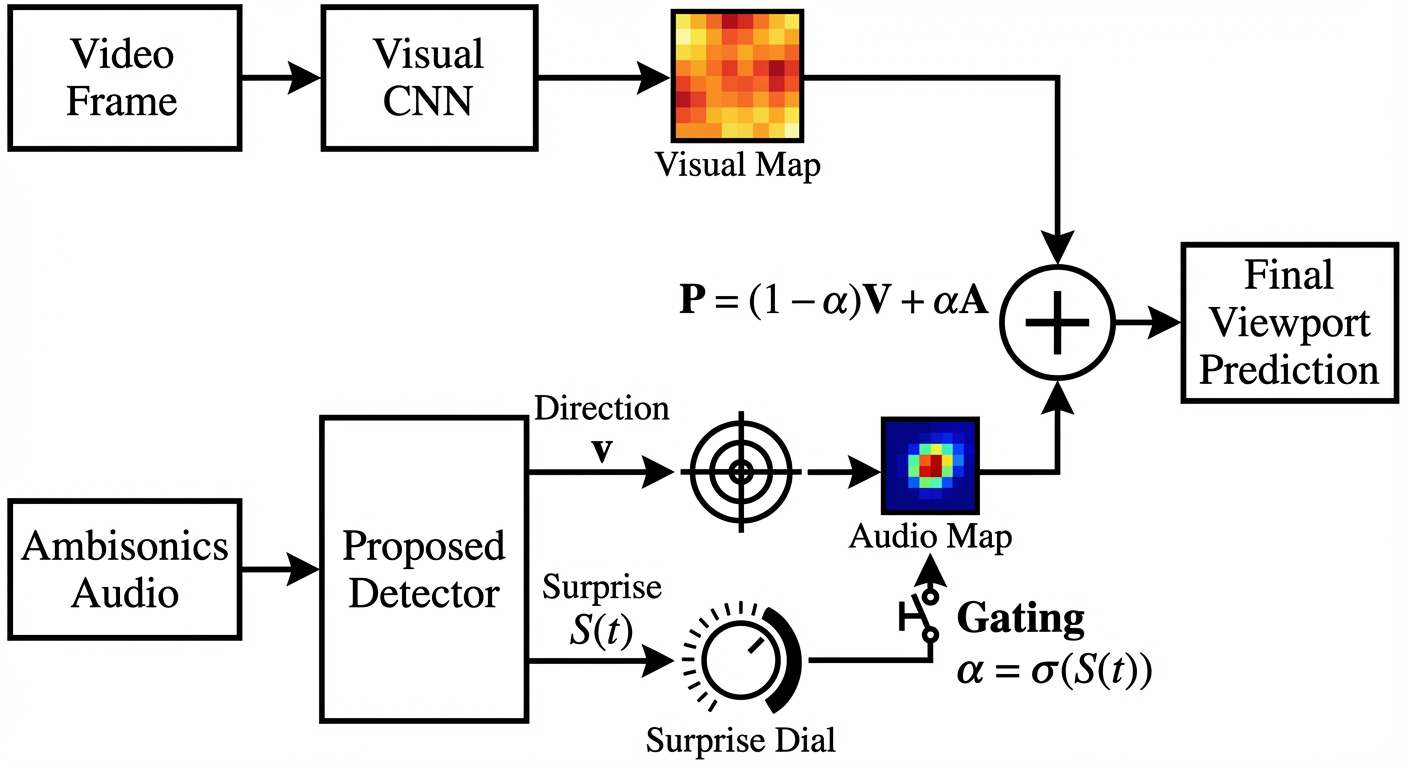}
\caption{\textbf{Audio-Visual Fusion Module.} The system operates as a gated ensemble. The Visual Saliency Net proposes a standard heatmap $\mathbf{H}_v$. The Audio Encoder outputs a directional vector $\mathbf{v}_a$ and a scalar surprise $S(t)$. The surprise score is mapped to a mixing coefficient $\alpha \in [0, 1]$ via a sigmoid gate, blending the visual heatmap with the audio cue to form the final viewport probability $\mathbf{P}_{final}$.}
\label{fig:fusion}
\end{figure}
\subsection{Detection of Audio-Triggered Gaze Shifts}
\begin{figure*}[t]
\centering
\includegraphics[width=0.8\linewidth]{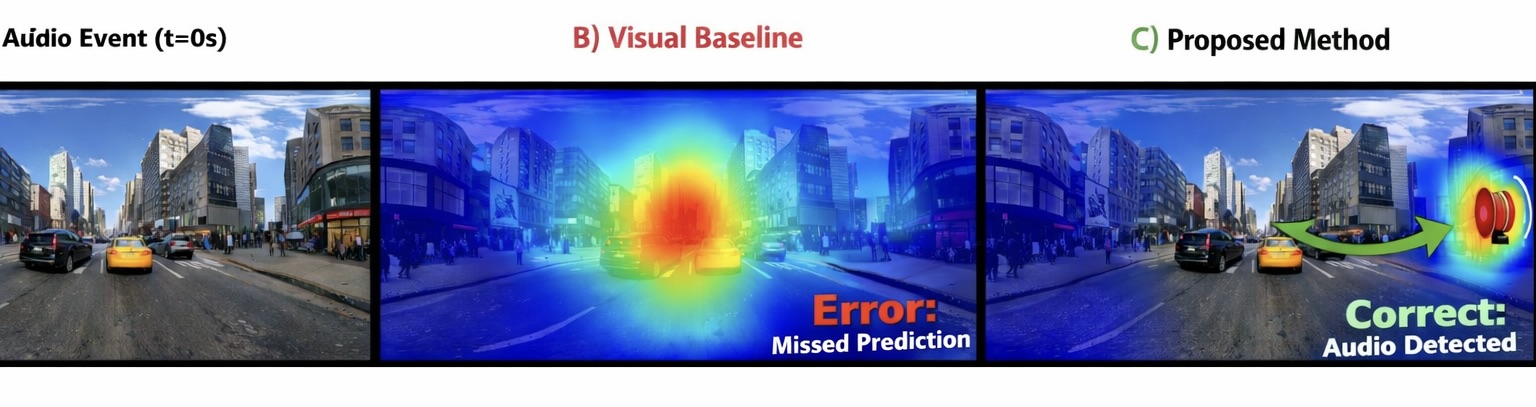}
\caption{\textbf{Qualitative comparison during an off-screen auditory event.} The user is watching a street scene. (A) \textbf{Ground Truth:} An ambulance siren (red icon) activates behind the user (unseen). (B) \textbf{Visual Saliency Baseline:} The model predicts attention remains on the visible street center, failing to anticipate the turn. (C) \textbf{Proposed Method:} The audio surprise signal modifies the prediction heat map, correctly identifying the rear region as the target for the upcoming saccade.}
\label{fig:qualitative}
\end{figure*}
The critical failure mode for existing VP systems is an ``Audio-Triggered Gaze Shift''---a rapid head rotation toward an off-screen sound source.
We algorithmically isolated 450 such events from the dataset using an objective threshold of $>30^\circ$/s angular head velocity to test the models under stress.
Table~\ref{tab:accuracy} details the prediction accuracy, measured by the Tile Overlap Ratio (TOR) between the predicted and actual viewports.
\begin{table}[h]
\caption{Viewport Prediction Accuracy (Tile Overlap Ratio)}
\label{tab:accuracy}
\centering
\begin{tabular}{l|c|c}
\hline
\textbf{Method} & \textbf{General Scenarios} & \textbf{Surprising Events*} \\
& (Mean $\pm$ Std) & (Mean $\pm$ Std) \\
\hline
Trajectory-Only (Inertia) & $0.91 \pm 0.04$ & $0.23 \pm 0.11$ \\
Visual Saliency & $\mathbf{0.93} \pm 0.03$ & $0.46 \pm 0.15$ \\
\textbf{Proposed (Hybrid)} & $0.92 \pm 0.03$ & $\mathbf{0.87} \pm 0.08$ \\
\hline
\multicolumn{3}{l}{\textit{*Events where audio source is outside the current Field of View.}}
\end{tabular}
\end{table}
\textbf{Analysis:}
All methods perform comparably in general scenarios (TOR $> 0.9$), but baselines collapse under surprising events.
The Trajectory model ($0.23$) assumes continued forward gaze; the Visual model ($0.46$) cannot react to off-screen sounds until the turn has begun.
The Proposed method maintains high accuracy ($0.87$) by detecting the surprise signal ${\sim}200$ms before head rotation begins ($p < 0.01$, paired t-test).

\subsection{Streaming Efficiency and QoE Impact}
To quantify the system-level impact, we simulated a tile-based MPEG-DASH adaptive streaming environment (4K resolution, $4 \times 8$ tiling), where surprise detection runs on the server/edge during ingestion rather than on the client HMD.
We measured the \textbf{Wasted Bandwidth Ratio} (data downloaded for tiles never seen) and the \textbf{Structural Similarity Index} (SSIM) of the viewed content.
\begin{table}[h]
\caption{Streaming Efficiency and QoE Metrics}
\label{tab:qoe}
\centering
\begin{tabular}{l|c|c|c}
\hline
\textbf{Method} & \textbf{Bitrate} & \textbf{Wasted BW} & \textbf{SSIM} \\
& (Mbps) & (\%) & (Mean) \\
\hline
Trajectory-Only & 14.5 & 44.8\% & 0.82 \\
Visual Saliency & 15.2 & 38.2\% & 0.86 \\
\textbf{Proposed} & \textbf{12.4} & \textbf{21.8\%} & \textbf{0.94} \\
\hline
\textit{Improvement} & \textit{-18.4\%} & \textit{-42.9\%} & \textit{+9.3\%} \\
\hline
\end{tabular}
\end{table}
The results in Table~\ref{tab:qoe} demonstrate a substantial efficiency gain.
By correctly predicting off-screen events, the Proposed method avoids pre-fetching high-quality tiles for the wrong direction.
This results in an $18.4\%$ reduction in overall bitrate consumption while simultaneously improving the Structural Similarity Index (SSIM) of the viewed content from 0.86 to 0.94.
The ``Wasted Bandwidth'' drops dramatically ($44.8\% \rightarrow 21.8\%$) because the system effectively ``hears'' where to download next, rather than guessing blindly.

In conclusion, the experimental results validate that while visual and inertial cues are sufficient for steady-state viewing, audio surprise modeling is indispensable for handling the dynamic, attention-grabbing events that define immersive experiences.

\section{Conclusion and Future Directions}
\label{sec:conclusion}
This paper presented a self-learning framework for detecting surprising auditory events in spatial audio for viewport prediction in 360-video streaming.
The architecture combines SE(3)-equivariant graph neural network encoding of First-Order Ambisonics audio with recurrent temporal modeling, producing a spatially and temporally aware surprise signal.
Surprise is formulated as prediction error, naturally capturing temporal attention decay — events are most surprising at onset and diminish as the listener adapts.
Experiments on AVTrack360 demonstrate that integrating audio surprise with visual saliency reduces bitrate waste and improves QoE compared to visual-only baselines.

Several directions merit further investigation.
\textbf{Multi-Modal Fusion:} The current work treats audio and visual saliency as separate signals; joint audio-visual models that capture cross-modal interactions could yield more robust predictions.
\textbf{Semantic Enrichment:} Combining statistical surprise with semantic audio understanding (e.g., speech recognition, event classification) could distinguish between ``novel'' and ``important'' sounds.
\def\IEEEbibitemsep{0pt plus 0.5pt}
\bibliographystyle{IEEEbib}
\bibliography{icme2026references}

\end{document}